\documentstyle[preprint,prb,aps]{revtex}

\begin{document}


\title{Deviations from Matthiessen's Rule for ${\rm SrRuO_3}$ and ${\rm CaRuO_3}$ }
\author{L. Klein, Y. Kats, and N. Wiser}
\address{
Department of Physics, Bar-Ilan University, Ramat-Gan 52900, Israel }
\author{M. Konczykowski}
\address{
Laboratoire des Solides Irradies, Ecole Polytechnique, 91128 Palaiseau
Cedex, France}
\author{J. W. Reiner, T. H. Geballe, M. R. Beasley, and A. Kapitulnik}
\address{
Edward L. Ginzton Laboratories, Stanford University, Stanford, California 94305}
\date{\today}

\maketitle

\begin{abstract}
We have measured the change in the resistivity of thin films 
of ${\rm SrRuO_3}$ and ${\rm CaRuO_3}$ upon introducing point defects by electron
irradiation at low temperatures, and we find significant
deviations from Matthiessen's rule. For a fixed irradiation dose, the
induced change in resistivity {\it decreases} with increasing temperature.
Moreover,  for a fixed temperature, the increase in resistivity with
irradiation is found to be {\it sublinear}. We suggest that the observed
behavior is due to the marked anisotropic scattering of the electrons 
together with their relatively short mean free path (both characteristic of
many metallic oxides including cuprates)  which amplify effects related to the
Pippard ineffectiveness condition.
\end{abstract}



Deviations from Matthiessen's rule (DMR) in a metal provide unique
microscopic insight into the electron scattering processes. 
Here, we have used electron irradiation to study the DMR in the
ruthenium-based perovskites ${\rm SrRuO_{3}}$ and ${\rm CaRuO_{3}}$. These
ruthenates have been studied extensively in recent years,\cite
{allen,mazin,santi,sro,cro1,cro2} with various experimental results indicating
that these are strongly correlated systems, similar in many ways to the high-%
$T_{c}$ cuprates. Therefore, there is special interest in exploring the DMR
in these systems.

Usually, the DMR
have been studied by comparing the electrical resistivities of samples
containing different concentrations of impurities.\cite{DMR} However, this
method may induce changes in the resistivity that are not related to the
concentration of point defects, thus limiting the possibility of a rigorous
and sensitive study of the DMR. On the other hand, it has been pointed out 
\cite{millis} that examining the resistivity change of a sample whose
residual resistivity is moderately increased by electron irradiation would
allow a detailed study of the DMR which could be compared with theoretical
predictions.

${\rm SrRuO_{3}}$ and ${\rm CaRuO_{3}}$ are pseudo-cubic
perovskites having similar electronic structure and comparable
resistivities. Nevertheless, the slight difference in their lattice
parameters leads to important differences in their magnetic behavior. ${\rm %
SrRuO_{3}}$ is an itinerant ferromagnet with $T_{c}\sim 150{\rm \ K}$ (in
films), whereas ${\rm CaRuO_{3}}$ does not exhibit long-range order down to
at least $1.8{\rm \ K}$. Since ${\rm SrRuO_{3}}$ and ${\rm CaRuO_{3}}$ are
very similar except for magnetic ordering, studying both systems allows us
to determine whether the observed DMR are due to the magnetic contribution.

Our measurements show significant DMR for both ${\rm SrRuO_{3}}$ and ${\rm %
CaRuO_{3}}$. The experimental quantity of interest is the change in
resistivity ($\rho $) upon electron irradiation, $\Delta \rho
_{irr}(D,T)=\rho (D,T)-\rho (0,T)$, where $D$ is the irradiation dosage and $%
T$ is the temperature. We measured $\Delta \rho _{irr}(D,T)$ as a function
of $D$ for fixed $T$ and as a function of $T$ for fixed $D$. Our principal
results are the following: (i) for fixed dosage, $\Delta \rho _{irr}$ 
{\it decreases} with increasing temperature (see Fig.\,1), and (ii) for
fixed temperature, $\Delta \rho _{irr}$ increases {\it sublinearly} as
a function of dosage (see Fig.\,2). We explain these results in terms of
 anisotropic electron scattering (due to the anisotropic Fermi surface\cite
{allen,mazin,santi}) together with the Pippard ''ineffectiveness condition'' 
\cite{pipp1,pipp2,pipp3} for low-$q$ scattering due to a short mean free
path. Since anisotropic scattering and a short mean free path are also
characteristic of other metallic perovskites (including the cuprates), 
DMR may be expected to occur in these systems as well.

The samples used for electron irradiation were ${\rm SrRuO_{3}}$ films grown
on ${\rm SrTiO_{3}}$ and a ${\rm CaRuO_{3}}$ film grown on ${\rm NdGaO_{3}}$
(different substrates where chosen to minimize the lattice mismatch). The samples were
irradiated by a beam of 
 2.5 MeV electrons using the VINKAC set-up at
Ecole Polytechnique, composed of a Van de Graff accelerator and an
irradiation chamber connected to a closed-cycle hydrogen liquefier.
Electrons of such energy are known\cite{irr} to create point defects in
compounds similar to the perovskites used in our experiment. During
irradiation, the samples were immersed in liquid hydrogen ($20{\rm \ K}$) to
prevent diffusion and clustering of the generated point defects. In the
first warm-up, we observe a small decrease in $\Delta \rho _{irr}$ which
indicates some defect migration when the temperature is increased. By analogy
with other electron-irradiated systems (e.g., ${\rm YBa_{2}Cu_{3}O_{7}}$)
whose defects were extensively studied with a tunneling-electron microscope,
this decrease is due to partial annealing or the formation of small clusters
which are equivalent to point defects. The defects remained stable in
subsequent temperature cycles.

Prior to irradiation, the films were patterned to allow resistivity
measurements on pairs of neighboring segments. During irradiation, one
segment of each pair was covered with lead to protect it from being
irradiated. Therefore, we could determine the effect of irradiation by
comparing the resistivities of neighboring segments, with the
lead-covered segment serving as reference. 
The fact that the comparison
is made between segments that are identical except for
the irradiation, and the fact that the resistivity of
 the two segments was measured simultaneously (thus
avoiding errors due to small variations in the sample
 temperature at the same nominal temperature
settings during different measurement runs) are crucial for 
reliable and sensitive measurements of the small
 changes in resistivity due to irradiation.

The main results of this paper are presented in Figs. 1 and 2. Figure\,1
shows the temperature dependence of $\Delta \rho _{irr}$ at a fixed dosage for 
${\rm SrRuO_{3}}$ and ${\rm CaRuO_{3}}$. At low temperatures, $\Delta \rho _{irr}\sim $ $3{\rm %
\ \mu \Omega \ cm}$ for both samples, but as the temperature increases, $\Delta \rho _{irr}$
decreases significantly. The similarity in the behavior of $\Delta \rho _{irr}$
for ${\rm SrRuO_{3}}$ and ${\rm CaRuO_{3}}$
indicates that the observed DMR is not caused by magnetic ordering. However,  the
pronounced ''step'' in $\Delta \rho _{irr}$ near the Curie temperature of 
${\rm SrRuO_{3}}$ suggests that the magnetic ordering does affect $\Delta \rho _{irr}$.

Figure\,2 shows $\rho $ as a function of dosage for ${\rm %
SrRuO_{3}}$ at ${\rm 20\ K}$. Whenever the density of the added point
defects is proportional to the dosage,\cite{nl} one expects a linear increase
in $\Delta \rho _{irr}$. However, we find a {\it sublinear} dependence; the inset clearly
shows a monotonic decrease in $d\rho /dD$.

We suggest that the key to understanding these results is twofold: (i) the
scattering is very anisotropic for the ruthenates; therefore, even
small-angle scattering can produce large changes in the velocity of the
electron, and (ii) the anisotropic scattering amplifies the Pippard
''ineffectiveness condition''\cite{pipp1} which states that the low-$q$ components of
scattering potentials are ineffective in scattering when $1/q$ is larger
than the electron mean free path.

The physics behind the ineffectiveness condition is straightforward. A
finite mean free path $\lambda $ implies that the $k$-vector of the electron
has an uncertainty of magnitude $\Delta k\sim 1/\lambda $. Therefore, if the
electron $k$-vector changes upon scattering by less than $\Delta k$, 
then 
the initial and final $k$-vectors are the same within their uncertainty. This is equivalent to the
electron not having been scattered at all.

The change in the electron $k$-vector upon scattering is given by the $q$%
-vector of the scattering potential, $V_{q}$. The ineffectiveness
condition implies that the mean free path $\lambda $ sets a lower bound $%
q_{min}=2\pi /\lambda $, so that $V_{q}$ with $q<q_{min}$ is excluded as a
scattering source.

The ineffectiveness condition explains the behavior of $\Delta \rho
_{irr}(D,T)$, both as a function of temperature (Fig. 1) and as a function
of irradiation dosage (Fig. 2).

Two different contributions determine the temperature-dependent decrease
of $\Delta \rho _{irr}$. First,  because $q_{min}{}$ increases as a function of
temperature, the resistivity due to the added
defects becomes smaller. Second, the irradiation-induced shortening of the mean free
path makes more small-$q$ phonons and magnons ''ineffective'', thus yielding
a negative contribution to $\Delta \rho _{irr}$.

Similarly, the sublinear increase of $\Delta \rho _{irr}$ with dosage is explained
by the decrease of  
 the contribution  {\it per defect} to $\rho $ due to
 the progressive increase in $q_{min}{}$. 

Why are similar DMR effects not observed in $all$ metals? For metals having a nearly-spherical
Fermi surface, the small-$q$ components of the scattering potential
contribute very little to the resistivity due to the ($1-\cos \theta $)
''transport factor'' ($\theta $ is the scattering angle). The approximate
expression $(1-\cos \theta )\propto q^{2}$ for the transport factor reflects
the fact that in simple metals, small scattering angles imply small changes in
electron velocity (hence, current). On the other hand, for a material with
an anisotropic Fermi surface, even small-angle scattering can lead to a {\it %
large} change in the electron velocity. As a result, the transport
factor is of order unity.\cite{unity} 

The effect of a non-spherical Fermi surface has been calculated, for
instance, for aluminium.\cite{alum} Using the correct
transport factor, which turns out to be of order unity, accounts for the
observed low-temperature $T^{3}$ dependence for $\rho (T)$, rather than the
usual $T^{5}$ dependence.

For the materials studied here, $d$-electrons are mainly responsible for the
current and the electron scattering is very anisotropic. Therefore, similar
to aluminium, the transport factor is expected to be of order unity over
much of the non-spherical Fermi surface. 
Moreover, these materials are characterized by a short mean free path which
implies large $q_{min}$.

We now consider the {\it magnitude} of $\Delta \rho _{irr}$. The
experimental results for ${\rm SrRuO_{3}}$ and ${\rm CaRuO_{3}}$ are
 similar, and we shall concentrate on the data for ${\rm %
SrRuO_{3}}.$\ According to Fig.\,1a, as the temperature increases to $100{\rm %
\ K}$, the value of $\Delta \rho _{irr}$ decreases from $3.0 {\rm \ \mu \Omega
\,cm}$ (its low-temperature value) to $1.4{\rm \ \mu \Omega \,cm}$. This $1.6%
{\rm \ \mu \Omega \,cm}$ decrease in $\Delta \rho _{irr}$ is the number we
wish to calculate.

An explicit resistivity calculation requires the scattering matrix
elements and the complete phonon and magnon spectrum, and performing
 integrals over the anisotropic Fermi surface. None of
these ingredients are known with sufficient precision. Therefore, we shall
adopt the more modest goal of carrying out a simplified resistivity
calculation to see whether we obtain reasonable
agreement with experiment. To that end, we use the Bloch-Gruneisen
approximation for both the electron-phonon and the electron-magnon
resistivity. Although such a calculation will not yield
quantitatively reliable numbers, we are here looking for a qualitative
explanation of the behavior of $\Delta \rho _{irr}$, rather than
attempting a first-principles calculation.

Following the textbooks,\cite{Ziman 364} we write the Bloch-Gruneisen
resistivity as follows: 
\begin{equation}
\rho =C\int\nolimits_{q_{min}(D,T)}^{Q}dq\,q\,|V(q)|^{2}\,F[\hbar \omega
(q)/k_{B}T]  \label{Eq. 1}
 \\
\text{;       } \ \ \ \ \ F(z)=\frac{z^{2}}{(e^{z}-1)(1-e^{-z})}  \nonumber
\end{equation}
\newline
where $q_{min}(D,T)$ depends on both the dosage $D$ and the temperature
through its dependence on the resistivity, and the upper limit is the Debye
wavenumber $Q$. At low $q$ (the region of interest here), $\omega (q)\propto
q$ for phonons, whereas for magnons, $\omega (q)\propto q^{2}$. The Debye
temperature and some other parameters are different, of course, for phonons
and for magnons. For the form factor $V(q)$, which describes the scattering
of an electron by a unit cell, we used the $q$-dependence which has been
calculated\cite{Harrison 35} for the simple metals.  We used the kinetic theory
of metals\cite{kinetic} to relate $\rho $ to $\lambda $ for the calculation
of $q_{min}=2\pi /\lambda $.

Equation (\ref{Eq. 1}) differs from the textbook expression in two respects.
First, we have approximated the transport factor by unity. Second, the
lower limit on the integral is $q_{min}(D,T)$, rather than zero, which takes
into account the ineffectiveness condition.

The resistivity of the defects is also given by Eq. (\ref{Eq. 1}), if one
omits the function $F(z)$ and extends the range of integration to twice the
Fermi momentum, which can be estimated from the calculated\cite{santi} Fermi
surface area.

Performing the calculation according to Eq. (\ref{Eq. 1}) yields a decrease
in $\Delta \rho _{irr}$ of $1.1{\rm \ \mu \Omega \,cm}$ at $100{\rm \ K}$.
Half of this value results from the decrease
in the resistivity of the added
defects from its zero-temperature value, and the other half results from the
decrease in the phonon and magnon resistivities due to the irradiation-induced
increase of $q_{min}$.\ The calculated value of $1.1{\rm \ \mu \Omega \,cm}$
is remarkably close to the
experimental value of $1.6{\rm \ \mu \Omega \,cm}$, in view of the crudity
of the calculation, and should not be taken seriously as a quantitative
result. However, it demonstrates that the ineffectiveness condition does
indeed account for the {\it magnitude} of the observed decrease in $\Delta
\rho _{irr}$ with temperature. 

At high temperatures our semi-classical treatment of $\rho _{irr}$ is 
not applicable, 
 since there are clear indications for the breakdown of Boltzmann 
transport theory in this regime.\cite{allen,sro,cro1}
Therefore, a different theoretical approach is required.

A  puzzling feature of the temperature dependence of $\Delta \rho _{irr}$ that still needs to be
explained is the sharp ''step'' in
$\Delta
\rho _{irr}$ of ${\rm SrRuO_{3}}$ near $T_{c}\sim 153{\rm \ K}$ (see Fig.\,1a). While one can
attribute a decrease in $\Delta \rho _{irr}$ near $T_{c}$ to the rapid increase of the
magnetic resistivity, the sharp ''step'' is surprising because
the magnetization approaches zero {\it gradually}, as the temperature
approaches $T_{c}$ from below. Furthermore, when  comparing the resistivity of the irradiated and
unirradiated segments in magnetic field (see Fig.\,3), additional unexpected behavior is observed near
$T_c$. Attributing the ''step'' in $\Delta \rho _{irr}$
to the decrease in the magnetization, the sharp ''step'' 
 should be smeared out by the
field. 
 This was indeed observed, as shown in the figure, but we also found the
unexpected result that below $T_{c}$, the application of the field $reduces$  $\Delta \rho _{irr}$. 

Both anomalies are resolved by assuming that the irradiation slightly
decreases $T_{c}$ by about $\Delta T_c \sim 0.4{\rm \ K}$, which rescales the temperature dependence
of the magnetic resistivity. 
Denoting the magnetic resistivity before  and after 
irradiation by $\rho_m$ and $\rho_m'$, respectively, we have 
 near $T_c$ the relation
 $\rho_m'(T,H) \simeq \rho_m(T+\Delta T_c,H)$. This change in 
the magnetic resistivity  makes a
temperature dependent contribution to  $\Delta \rho _{irr}$ that can be evaluated 
by determining $d\rho_m / dT$.\cite{Tc}  Subtracting this contribution from 
$\Delta \rho _{irr}$ yields the results shown in the upper
inset of Fig.\,3. These results indicate that below $T_{c}$, as expected, the
field does not reduce $\Delta \rho _{irr}$, and that the decrease of $\Delta
\rho _{irr}$ below $T_{c}$ is rather smooth.

The assumption that $T_c$ decreased by $\Delta T_c \sim 0.4{\rm \ K}$ due to irradiation
is supported by the observed relation between $T_c$ and the residual resistivity in our unirradiated
films. 
Furthermore, if we measure the magnetoresistance of the irradiated and unirradiated 
segments as a function of temperature, we see that the two curves coincide almost exactly if we shift one
relative to the other by $\sim 0.4{\rm \ K}$ (see lower inset of Figure 3).

Turning now to Fig.\,2, the inset presents the data for $d\rho /dD$,
 which gives a quantitative measure of
the sublinearity of $\Delta \rho _{irr}$. As the resistivity increases upon irradiation from $18.4$ to
$23.1  {\rm \ \mu \Omega \,cm}$, the
value of $d\rho /dD$ is seen to decrease by $15\%$. This is the number 
we wish to calculate.

Differentiating the resistivity integral with respect to $D$ for each
contribution to the resistivity (phonons do not contribute at low
temperatures) yields $3\%$ for the decrease in $d\rho /dD$ upon irradiation.
This value is much smaller than the experimental value of $15\%$. However,
the resistivity integral in (\ref{Eq. 1}) assumes isotropic electron
scattering. If the marked anisotropy of the scattering over the Fermi
surface is approximately taken into account, numerical studies indicate that
the calculated value of $d\rho /dD$ increases by a factor of $2$-$3$. (At
the higher temperature of $100{\rm \ K}$, the effect is only marginal,
because larger values of $q_{min}$ are involved.) Thus, a more realistic
calculation of $d\rho /dD$ would yield about half the observed
 sublinearity of $\Delta \rho _{irr}$. 

In conclusion, we have clearly demonstrated that ${\rm SrRuO_3}$ and ${\rm CaRuO_3}$
exhibit significant DMR. We suggest that this behavior stems from the
Pippard ineffectiveness condition, whose effect on the resistivity of these
materials is amplified by their anisotropic scattering and short mean free
path. We believe that these results will contribute to a 
better understanding of the transport properties of the 
compounds studied, as well as other similar systems.

We thank J. S. Dodge for useful discussions. This research was supported by
The Israel Science Foundation founded by the Israel Academy of Sciences and
Humanities and by Grant No. 97-00428/1 from the United States-Israel
Binational Science Foundation (BSF), Jerusalem, Israel.

\begin{figure}[tbp]
\caption{ Resistivity change due to irradiation ($\Delta \protect\rho_{irr}$%
) for thin films of (a) ${\rm SrRuO_3}$ and (b) ${\rm CaRuO_3}$ as a
function of temperature. The dosage for both samples is about $5\times
10^{18} {\rm \ electrons/cm^2}$. The insets show the resistivity of the
films before irradiation.}
\end{figure}

\begin{figure}[tbp]
\caption{{\it In situ} resistivity ($\protect\rho $) of a ${\rm SrRuO_{3}}$
film during irradiation as a function of dosage ($D$). During irradiation,
the sample was immersed in liquid hydrogen.
 The inset shows the derivative of the resistivity with
respect to the dosage ($d\protect\rho /dD$) as a function of dosage.}
\end{figure}

\begin{figure}[tbp]
\caption{ Resistivity change due to irradiation ($\Delta \protect\rho_{irr}$%
) of a ${\rm SrRuO_3}$ film in zero field (open circles) and in a magnetic
field of ${\rm 1 \ T}$ (full circles). The lower inset shows the
magnetoresistance ($ {\rm MR} =\rho(H)-\rho(0)$) of the unirradiated (full circles) and the irradiated
segments (crosses). The upper inset shows $\Delta \protect\rho^*_{irr}$
which is $\Delta \protect\rho_{irr}$ with the change in $T_c$ (see text)
taken into account. }
\end{figure}


\begin{references}
\bibitem{nl}  This is true as long as the density of defects is not too
large and the probability of clustering of defects is small.
\bibitem{Tc}  The behavior of $d\rho_m / dT$ near $T_c$ is 
determined by subtracting the derivative of the
{\it non-magnetic} components from the derivative of the
total resistivity  ($d\rho / dT$).  The derivative of the
{\it non-magnetic} components is found by extrapolation from
temperatures high enough above $T_c$ that
$\rho_m$ is constant.
\end{references}
\end{document}